\documentstyle [12pt] {article}
\begin{document}

\def\fbpi{f_+^{B\pi}}
\def\fdpi{f_+^{D\pi}}
\def\be{\begin{eqnarray}}
\def\en{\end{eqnarray}}
\def\non{\nonumber}
\def\la{\langle}
\def\ra{\rangle}
\def\pr{{\sl Phys. Rev.}~}
\def\prl{{\sl Phys. Rev. Lett.}~}
\def\pl{{\sl Phys. Lett.}~}
\def\np{{\sl Nucl. Phys.}~}
\def\zp{{\sl Z. Phys.}~}

\font\el=cmbx10 scaled \magstep2
{\obeylines
\hfill IP-ASTP-10-94
\hfill May, 1994}

\vskip 1.5 cm

\centerline{\el Form Factors for $B\to\pi$ and $D\to\pi$ Transitions}
\medskip
\bigskip
\medskip
\centerline{\bf Hai-Yang Cheng}
\medskip
\centerline{ Institute of Physics, Academia Sinica}
\centerline{Taipei, Taiwan 11529, Republic of China}
\bigskip
\bigskip
\bigskip
\centerline{\bf Abstract}
\bigskip
  Armed
with the information on the form factor $\fdpi(0)$ inferred from recent CLEO
measurements of SU(3)-breaking effects in charmed meson decays, we have
studied the form factor $\fbpi$. In the heavy quark limit, $\fbpi(q^2)$
is related to $\fdpi(q^2)$ in the kinematic region close to zero recoil.
Assuming pole dominance for its $q^2$ dependence, $\fbpi(0)$ is estimated to
be $\approx 0.39\,.$ If the
requirement of heavy quark symmetry is relaxed so that it applies only to
soft pion emissions from the heavy meson, we find that $\fbpi(0)$ is more
likely of order $0.55\sim 0.60\,.$

\pagebreak

A reliable determination of the quark mixing matrix element $V_{ub}$ from the
semileptonic decay mode $\bar{B}\to\pi\ell\bar{\nu}$ ($\ell=e,~\mu$) requires
a knowledge of the $\bar{B}\to\pi$ transition form factor $f_+^{B\pi}$ at
$q^2=0$. In the past, this form
factor has been calculated using the nonrelativistic quark model [1], QCD sum
rules [2], and heavy quark symmetry in synthesis with chiral symmetry [3-5].
A systematic analysis of the $1/m_b$ correction to the weak form factors
$f_\pm^{B\pi}$ was recently studied in the framework of the heavy quark
effective theory [6]. In the method of heavy quark symmetry, form factors
$f_\pm^{B\pi}$ can be related in a model-independent way to the form factors
$f_\pm^{D\pi}$ in the kinematic region close to zero recoil. In order to
extract $f_+^{B\pi}(0)$ from the available experimental information of
$f_+^{D\pi}(0)$, an extrapolation of the form factors from zero recoil to
maximum recoil (i.e. $q^2=0$) has to be assumed. However, unlike the well
measured form factor $f_+^{DK}(0)$, the present experimental data on
$f_+^{D\pi}(0)$ are still plagued with large statistic
and systematic errors. Fortunately, this situation was changed recently, Two
new measured SU(3)-breaking effects in charm decays to be discussed
later are very sensitive to the relative magnitude of the form factors
$f_+^{D\pi}(0)$ and $f_+^{DK}(0)$. By fitting to the data, we found a best
fit of $f_+^{D\pi}(0)/f_+^{DK}(0)$, and hence $f_+^{D\pi}(0)$. In this paper,
we will study the form factor $\fbpi(0)$ in two stages. In the first stage,
heavy quark symmetry is applied so that $\fbpi(q^2)$ is related to
$\fdpi(q^2)$ near zero recoil. In the second stage, the requirement of heavy
quark symmetry is relaxed, namely it applies only to soft pion emissions from
the heavy meson. We then make comments.

    The matrix element of the $\bar{B}\to\pi$ transition is usually
parametrized as
\be
\la\pi(p_\pi)|\bar{q}\gamma_\mu b|\bar{B}(v)\ra =\,f_+^{B\pi}(q^2)(m_Bv+p_\pi
)_\mu+f_-^{B\pi}(q^2)(m_Bv-p_\pi)_\mu,
\en
or equivalently,
\be
\la\pi(p_\pi)|\bar{q}\gamma_\mu b|\bar{B}(v)\ra =\,  f_1^{B\pi}(q^2)(m_Bv+
p_\pi)_\mu+{m_B^2-m_\pi^2\over q^2}q_\mu[f_0^{B\pi}(q^2)-f_1^{B\pi}(q^2)],
\en
where $q=m_Bv-p_\pi$, and the form factors $f_\pm^{B\pi}$ and $f_{0,1}^{B\pi}$
are related by
\be
f_1^{B\pi}(q^2)=\,\fbpi(q^2),~~~f_0^{B\pi}(q^2)=\,\fbpi(q^2)+{q^2\over m_B^2
-m_\pi^2}f_-^{B\pi}(q^2).
\en
To avoid unphysical poles at $q^2=0$ in Eq.(2), one must have $f_0(0)=f_1(0)$.
In the $m_b\to \infty$ limit, the matrix element $\la\pi|\bar{q}\gamma_\mu
h_b|\bar{B}\ra$ scales as $\sqrt{m_B}$, where $h_b$ is the velocity-dependent
effective heavy quark field for the $b$ quark. Since
\be
\bar{q}\gamma_\mu b=\,c(\mu)\bar{q}\gamma_\mu h_b(\mu)
\en
at the subtraction scale $\mu<m_b$, and the large logarithmic contribution to
$c(\mu)$ has been evaluated in Ref.[7], it follows from Eqs.(1) and (2) that
\be
(f_++f_-)^{B\pi}(q_B^2) &=& C_{bc}\sqrt{ m_D\over m_B}(f_++f_-)^{D\pi}(q_D^2),
\non  \\  (f_+-f_-)^{B\pi}(q_B^2) &=& C_{bc}\sqrt{ m_B\over m_D}(f_+-f_-)
^{D\pi}(q_D^2),   \\    C_{bc} &=& \left({\alpha_s(m_b)\over \alpha_s(m_c)}
\right)^{-6/25}, \non
\en
where $q^2_B=(m_Bv-q)^2,~q^2_D=(m_Dv-q)^2$. It is easily seen that
the relations (5), which are first derived in Ref.[8], are valid provided
that $p$ does not scale with $m_{c,b}$ or $v\cdot p<< m_{c,b}$. Eqs.(5) lead
to
\be
f_+^{B\pi}(q_B^2) &=& {C_{bc}\over 2\sqrt{m_Bm_D}}\,[(m_B+m_D)f_+^{D\pi}
(q_D^2)-(m_B-m_D)f^{D\pi}_-(q_D^2)],   \\ f_-^{B\pi}(q_B^2) &=& {C_{bc}\over
2\sqrt{m_Bm_D}}\,[-(m_B-m_D)f_+^{D\pi}(q_D^2)+(m_B+m_D)f^{D\pi}_-(q_D^2)].
\en
Note that since $(f_++f_-)^{B\pi}$ scales as $1/\sqrt{m_B}$ and $(f_++f_-)^
{D\pi}$ as $1/\sqrt{m_D}$, Eqs.(6) and (7) are sometimes further reduced to
\be
\fbpi(q_B^2)\approx -f_-^{B\pi}(q_B^2)=\,C_{bc}\sqrt{ m_B\over m_D}f_+^{D
\pi}(q_D^2),
\en
so that $\fbpi$ is expressed solely in terms of the physically measurable
quantity $\fdpi$, where use of the heavy-quark-symmetry approximation
$f_-^{D\pi}(q_D^2)\approx -\fdpi(q^2_D)$ has been made.

     Recently it has been demonstrated that the
form factor $f_1(q^2)$ has a monopole behavior in the combined large $N_c$,
heavy quark and chiral limits [5]:
\be
f_{1}(q^2)=\,{f_{1}(0)\over 1-{q^2\over m^2_{1}}},
\en
where $m_1$ is the mass of the lowest-lying $1^-$ resonance that couples to
the weak current. Such a behavior is also seen in many QCD
sum rule calculations [2]. However, this single pole behavior does not hold
for the form factor $f_0(q^2)$, as one can see from Eq.(3) that
\be
f_0^{B\pi}(q^2)\approx \fbpi(q^2)\left(1-{q^2\over m_B^2-m^2_\pi}\right)
\en
under the heavy-quark-symmetry relation $f_-^{B\pi}(q^2)\approx -\fbpi
(q^2)$. Hence, the $q^2$ dependence of $f_0$ is different from that of
$f_+$ by an additional pole factor [9]. In fact, if we follow Ref.[1] to
assume that $f_{0}(q^2)=f_{0}(0)/[1-(q^2/ m^2_0)]$
with $m_0$ being the $0^+$ pole mass, then it is easily seen that
\be
f_-^{B(D)\pi}(0)=\,\left(m^2_{B(D)\pi}-m^2_\pi\right)\left({1\over m^2_0}-
{1\over m_1^2}\right)f_+^{B(D)\pi}(0).
\en
Using $m_0=2.47~(5.99)$ GeV and $m_1=2.01~(5.32)$
GeV [1] for the form factors $f_{0,1}$ in $D-\pi$
($\bar{B}-\pi$) transition, we find from Eq.(11) that
\be
f_-^{B\pi}(0)=-0.21\fbpi(0),~~~~f_-^{D\pi}(0)=-0.29\fdpi(0),
\en
which are substantially different from heavy-qaurk-symmetry expectations.
{}From Eqs.(8) and (9) we find
\be
\fbpi(q_m^2) &=& 1.85\fdpi(q_m^2),   \\
\fbpi(0) &=& 0.47\fdpi(0),
\en
where $q_m^2$ is the momentum transfer squared at zero recoil and it is
understood to be $(m_B-m_\pi)^2$ for $\bar{B}-\pi$ transition and
$(m_D-m_\pi)^2$ for $D-\pi$ transition.

Presently, there are only two available experimental information on the form
factor $\fdpi(0)$. An earlier measurement
of the Cabibbo-suppressed decay $D^0\to \pi^-\ell^+\nu$ by Mark III yields
$\left|{f_+^{D\pi}(0)/ f_+^{DK}(0)}\right|=\,1.0^{+0.6}_{-0.3}\pm 0.1$ [10,11],
while a very recent CLEO-II measurement of $D^+\to \pi^0\ell^+\nu$ gives
$\left|{f_+^{D\pi}(0)/ f_+^{DK}(0)}\right|=\,1.29\pm 0.21\pm 0.11$ [12].
Though the latter perfers a larger $f_+^{D\pi}(0)$ over $f_+^{DK}(0)$,
its error is still very large. Fortunately, a better determination of the
ratio $f_+^{D\pi}(0)/ f_+^{DK}(0)$ can be inferred from the
recent CLEO measurements of the decay rates of $D^+\to\pi^+\pi^0$ and
$D^0\to K^+\pi^-$ [13], which give the values of the ratios
\be
R_1 &=& 2\left|{V_{cs}\over V_{cd}}\right|^2{\Gamma(D^+\to \pi^+\pi^0)\over
\Gamma(D^+\to\bar{K}^0\pi^+)}=\,3.29\pm 1.16\,,   \\
R_2 &=& \left|{V_{cs}^*V_{ud}\over V_{cd}^*V_{us}}\right|^2{\Gamma(D^0\to K^+
\pi^-)\over \Gamma(D^0\to K^-\pi^+)}=\,2.92\pm 0.95\pm 0.95\,.
\en
Naively both ratios are expected to be unity if SU(3) is a good symmetry. The
experimental values (15) and (16) thus appear quite striking at first glance.
We have shown recently that such large SU(3) violations in $R_1$ and $R_2$
can be accounted for by the accumulations of several small SU(3)-breaking
effects [14]. Crucial to our analysis is the relative magnitude of the form
factors $f_+^{D\pi}(0)>f_+^{DK}(0)$, a necessary ingredient for obtaining
the correct values of $R_1$ and $R_2$. A fit of the large-$N_c$ factorization
calculation [14] to $R_1$ yields
\footnote{A determination of the ratio $f_+^{D\pi}(0)/ f_+^{DK}(0)$ from
$R_2$ is contaminated by the presence of the $W$-exchange
diagrams and by possible final-state interactions.}
\be
f_+^{D\pi}(0)/ f_+^{DK}(0)\approx \,1.09\,.
\en
Using the average value [11]
\be
f_+^{DK}(0)=\,0.76\pm 0.02
\en
extracted from the recent measurements of $D\to K\ell\bar{\nu}$ by CLEO,
E687 and E691, we find
\footnote{What we have done here is opposite to the procedure in Ref.[14].
There, the value $\fdpi(0)\approx 0.83$ is first obtained by fitting the
factorization calculation to the measured decay rates of
$D^+\to\pi^+\pi^0$ [although this value is obtained in Ref.[14] by assuming
a monopole behavior for $f_0(q^2)$, the result remains the same if one
considers the $q^2$ dependence of $f_+(q^2)$ instead of $f_0(q^2)$.]
When combining with the experimental average value of
$f_+^{DK}(0)$ given by (18), it then implies Eq.(17). For a comparsion, the
values $f_+^{D\pi}(0)=0.69$ and $f_+^{DK}(0)=0.76$ are obtained in Ref.[1].
If these results were used in calculation, one would have obtained
$R_1=1.4\,$, in disagreement with data.}
\be
\fdpi(0)\approx 0.83\,.
\en
Substituting (19) into (14) yields
\footnote{Baur, Stech and Wirbel [1] obtained $\fbpi(0)=0.333\,$.}
\be
\fbpi(0)\approx 0.39\,.
\en

    Thus far we have determined the form factor $f_+^{B\pi}$ from
$f_+^{D\pi}$ via the heavy quark symmetry relation (8). It is known
that, within the framework of chiral perturbation theory which incorporates
both chiral and heavy quark symmetries [3,15], the form factor $\fbpi$ near
zero recoil is completely fixed by decay constants and the coupling constant
$g_{B^*B\pi}$. From the heavy-meson chiral perturbation theory given in
Refs.[3,15], we obtain ($f_\pi=132$ MeV)
\be
\fbpi(q_B^2)+f_-^{B\pi}(q_B^2)=\,{f_B\over f_\pi}
\en
in the small $v\cdot p_\pi$ limit. Owing to the near degeneracy of the
$B^*$ and $B$ masses,
it also becomes necessary to take into account the $B^*$ pole effects, which
are
\footnote{The sign of Eqs.(21-27) is opposite to that derived by M. Wise [3].}
\be
\fbpi(q_B^2)+f_-^{B\pi}(q_B^2) &=& -{f_{B^*}\over f_\pi}\,{gv\cdot p_\pi\over
v\cdot p_\pi+\Delta_B}\,\sqrt{m_B\over m_{B^*}},    \\
\fbpi(q_B^2)-f_-^{B\pi}(q_B^2) &=&\,{f_{B^*}\over f_\pi}\,{gm_{B^*}^2/m_B
\over v\cdot p_\pi+\Delta_B}\,\sqrt{m_B\over m_{B^*}}
\en
in the soft pion limit, where $\Delta_B=m_{B^*}-m_B$,
$g$ is the coupling constant in the heavy meson chiral Lagrangian [3,15],
and we have neglected terms of order $m_\pi/m_B$,
It follows from (22) and (23) that
\be
\fbpi(q_B^2)=\,{f_B\over 2f_\pi}\left( {f_{B^*}\over f_B}\,{gm_{B^*}\sqrt{
m_{B^*}/m_B}\over v\cdot p_\pi+\Delta_B}+1\right),  \\
f_-^{B\pi}(q_B^2)=\,-{f_B\over 2f_\pi}\left( {f_{B^*}\over f_B}\,{gm_{B^*}
\sqrt{m_{B^*}/m_B}\over v\cdot p_\pi+\Delta_B}-1\right).
\en
Likewise,
\be
\fdpi(q_D^2)=\,{f_D\over 2f_\pi}\left( {f_{D^*}\over f_D}\,{gm_{D^*}\sqrt{
m_{D^*}/m_D}\over v\cdot p_\pi+\Delta_D}+1\right),  \\
f_-^{D\pi}(q_D^2)=\,-{f_D\over 2f_\pi}\left( {f_{D^*}\over f_D}\,{gm_{D^*}
\sqrt{m_{D^*}/m_D}\over v\cdot p_\pi+\Delta_D}-1\right).
\en
It should be stressed that in the derivation of (24-27), heavy quark symmetry
has been applied only to the soft pion emissions from any
ground-state heavy meson so that the soft pion interaction with the heavy
meson is described by a single coupling constant $g$.
In the heavy quark limit where $\Delta_D=\Delta_B=0,~
f_{D^*}/f_D=1$, $f_{B^*}/f_B=1$, $f_B/f_D=C_{bc}\sqrt{m_D/m_B}$ [7], it
is easily seen that Eqs.(6) and (7) follow from Eqs.(24-27). Since
\be
v\cdot p_\pi+\Delta_B=\,{m_{B^*}\over 2}\left(1-{q_B^2\over m^2_{B^*}}\right),
\en
it is evident that, when $q_B^2$ is close to $q_m^2$, the form factor is
single pole dominated. It has been argued that [16], beyond the soft pion
limit, the relations (24-27) are still valid except that they must be
multiplied by a factor of $(1-\alpha v\cdot p_\pi/\Lambda_\chi)$, with
$\Lambda_\chi\sim 1$ GeV being a chiral symmetry breaking scale.
The fact that the pole behavior shown by Eq.(9) is seen in many
QCD sum rule calculations [2] over a large range of $q^2$ implies that
$\alpha\approx 0$.

   Unfortunately, since our present knowledge about the decay constants
and in particular the coupling constant $g$ is uncertain, we cannot predict
the form factors $\fbpi(0)$ and $\fdpi(0)$ reliably through the above
relations.
Nevertheless, we can still learn something about $\fbpi(0)$ based on the
aforementioned value of $\fdpi(0)$, lattice and QCD-sum-rule
calculations for decay constants. For the purpose of illustration, we will take
the central values of lattice calculations:
$f_B=187$ MeV [17], $f_D=208$ MeV [17], $f_{D^*}/f_D\approx 1.16f_{B^*}/f_B$
[18], and QCD sum rule result $f_{B^*}/f_B
\approx 1.1$ [19]. Assuming a
monopole behavior for the form factor $\fdpi(q^2)$, we find from Eq.(26)
\be
g=0.32\,,
\en
which is substantially smaller than what naively expected from the chiral
quark model [15]: $g=0.75\,$. Substituting (29) into (24) yields
\be
\fbpi(0)=\,0.53\,.
\en
One can see from Eqs.(24) and (26) that a smaller $\fbpi(0)$ requires a
smaller strong coupling $g$, and hence a larger ratio of $f_{D^*}/f_D$, which
is taken to be 1.3 in the above example. In general, the form factor
$\fbpi(0)$ is larger than $0.5\,$, to be compared with the value 0.39
inferred from Eq.(8) and the range $0.2\sim 0.3$ obtained in QCD sum rule
calculation [2].
As emphasized before, the chiral relations (24-27) are more general
than Eqs.(6,7) since the requirement of heavy quark symmetry is relaxed in the
former: it applies only to soft pion emissions from the heavy meson.
Thus we believe that $\fbpi(0)$ is more likely of order $0.55\sim
0.60\,$. This should be checked soon by lattice calculation.

   To conclude, using the value $\fdpi(0)\approx 0.83$ inferred from recent
CLEO measurements of SU(3) breaking effects in charm decays in conjunction with
experimental results for $f_+^{DK}(0)$, we have studied the
form factor $\fbpi(0)$. We find that it is $\approx 0.39$ in the heavy quark
limit and of order $0.55\sim 0.60$ when heavy quark symmetry is applied only
to soft-pion $B^*B\pi$ and $D^*D\pi$ couplings.

\vskip 2.0 cm
\centerline{\bf ACKNOWLEDGMENT}

    This work was supported in part by the National Science Council of ROC
under Contract No. NSC83-0208-M-001-014.

\pagebreak

\centerline{\bf References}
\medskip
\begin{enumerate}

\item M. Wirbel, B. Stech, and M. Bauer, \zp {\bf C29}, 637 (1985); N. Isgur,
D. Scora, B. Grinstein, and M.B. Wise, \pr {\bf D39}, 799 (1989).

\item C.A. Dominguez and N. Paver, \zp {\bf C41}, 217 (1988); A.A.
Ovchinnikov, {\sl Sov. J. Nucl. Phys.} {\bf 50}, 519 (1989); \pl {\bf B229},
127 (1989); V.L. Chernyak and I.R. Zhitnitski, \np {\bf B345}, 137 (1990);
P. Ball, V.M. Braun, and H.G. Dosch, \pl {\bf B273}, 316 (1991); \pr {\bf
D44}, 3567 (1991); S. Narison, \pl {\bf B283}, 384 (1992); P. Ball, \pr {\bf
D48}, 3190 (1993); V.M. Belyaev, A. Khodjamirian, and R. R\"uckl,
\zp {\bf C60}, 349 (1993); P. Colangelo, BARI-TH/93-152 (1993); P. Colangelo
and P. Santorelli, BARI-TH/93-163 (1993).

\item M.B. Wise, \pr {\bf D45}, 2188 (1992); G. Burdman and J. Donoghue, \pl
{\bf B280}, 287 (1992).

\item L. Wolfenstein, \pl {\bf B291}, 177 (1992); R.
Casalbuoni, A. Deandrea, N. Di Bartolomeo, R. Gatto, F. Feruglio, and
G. Nardulli, \pl {\bf B299}, 139 (1993).

\item B. Grinstein and P.F. Mende, Brown HET-928 (1993) and Brown HET-930
(1994).

\item G. Burdman, Z. Ligeti, M. Neubert, and Y. Nir, \pr {\bf D49}, 2331
(1994).

\item H.D. Politzer and M. Wise, \pl {\bf B206}, 681 (1988); {\it ibid}
{\bf B208}, 504 (1988).

\item N. Isgur and M. Wise, \pr {\bf D42}, 2388 (1990).

\item Q.P. Xu, \pl {\bf B306}, 363 (1993); M. Neubert and V. Rieckert, \np
{\bf B382}, 97 (1992).

\item Mark III Collaboration, J. Adler {\it et al.,} \prl {\bf 62}, 1821
(1989).

\item M. Witherell, talk presented at the XVI International Symposium
on Lepton-Photon Interactions, Ithaca, 10-15 August 1993.

\item CELO Collaboration, M.S. Alam {\it et al.,} \prl {\bf 71}, 1311 (1993).

\item CLEO Collaboration, M. Selen {\it et al.,} \prl {\bf 71}, 1973 (1993);
D. Cinabro {\it et al.,} \prl {\bf 72}, 406 (1994).

\item L.L. Chau and H.Y. Cheng, ITP-SB-93-49.

\item T.M. Yan, H.Y. Cheng, C.Y. Cheung, G.L. Lin, Y.C. Lin, and H.L. Yu,
\pr {\bf D46}, 1148 (1992).

\item L. Wolfenstein in Ref.[4].

\item C.W. Bernard, J.N. Labrenz, and A. Soni, \pr {\bf D49}, 2536 (1994).

\item A. Abada {\it et al.,} \np {\bf B376}, 172 (1992).

\item M. Neubert, \pr {\bf D46}, 1076 (1992).

\end{enumerate}

\end{document}